\documentclass{PoS}

\usepackage{pxfonts,amsfonts,wasysym}
\usepackage{units}
\usepackage{booktabs}
\usepackage{xspace}
\usepackage{graphicx}
\usepackage{subfig}	
\usepackage{wrapfig}	

\newcommand{\micron}{$\muup$m\xspace}

\title{Study of detection efficiency distribution and areal homogeneity of SiPMs}

\ShortTitle{Areal homogeneity test of SiPMs}

\author{\speaker{Michal Tesa\v{r}}, Christian Jendrysik, Jelena Ninkovi\'{c}, Frank Simon\\ 
        Max Planck Institute for Physics, Munich, Germany\\
        E-mail: \email{tesar@mpp.mpg.de}}

\abstract{
The analog hadron calorimeter for the International Linear Collider (ILC) of the CALICE collaboration utilized novel silicon detectors, the Sillicon Photomultipliers (SiPMs), for the detection of scintillation light coming from very small scintillator cells ($3 \times 3 \times 0.5 \;\mathrm{cm}^3$). This technology allows the construction of highly granular calorimeters used for excellent shower separation and therefore outstanding jet energy resolution using the particle flow concept. Since the SiPMs still have potential for further improvements, we developed a setup dedicated to the measurement of parameters like the areal distribution of relative photon detection efficiency and crosstalk probability which are used to characterize and compare different devices. Thanks to the precise positioning system together with excellent focusing of the light source we are capable of scanning whole SiPMs with an area of $\sim 1\;$mm$^2$ at a sub-pixel resolution of \unit[1-2]{\micron}. A further alignment compensation feature allows us to conduct separate analysis on each single pixel of a SiPM. We present the technical description of our measurement setup and results obtained for two types of SiPMs. The first one is a commercial Hamamatsu MPPC with \unit[50]{\micron} pixel size, the second one was the second iteration series SiMPl device developed and manufactured in the Semiconductor Laboratory of the Max Planck Institute.
}

\FullConference{International Workshop on New Photon-detectors,\\
		June 13-15, 2012\\
		LAL Orsay, France}

\begin{document}

\section{Introduction}
The future accelerator experiments, the Compact Linear Collider (CLIC) and International Linear Collider (ILC) intend to use particle flow algorithms \cite{flow}. These algorithms serve to conduct geometrical separation of charged and neutral particles with a help of the calorimeter and improve jet energy resolution this way. However, these algorithms require highly granular calorimeters, which must be equipped with silicon photomultipliers (SiPMs). For calorimeter design testing purposes, the CALICE collaboration has built a \unit[1]{m$^3$} analog hadron calorimeter prototype \cite{calice-ahcal}. This prototype used 7608 scintillator tiles with dimensions of $3 \times 3 \times 0.5 \;\mathrm{cm}^3$, each coupled to a SiPM. The analog hadron calorimeter of the International Large Detector (ILD) \cite{ild-loi} concept for ILC should have about 8~million channels. Project of that size requires mass production of well tested and understood SiPMs. 
It is also important to have a tool which is capable of characterizing, quantifying and comparing performance of different devices.

For this purpose, we developed a scanning setup, which measures the relative photon detection efficiency (PDE) and the crosstalk probability distribution across the SiPM area
with sub-pixel resolution. The main purpose of this test stand is to characterize SiPM prototypes produced in the semiconductor laboratory of the Max Planck Institute in Munich \cite{nin-hQE}. However, the setup gives us also opportunity to study and compare other manufacturers' devices.

\section{The Setup}

\begin{wrapfigure}{r}{0.48 \textwidth}
  \centering
    \includegraphics[width=0.50 \textwidth]{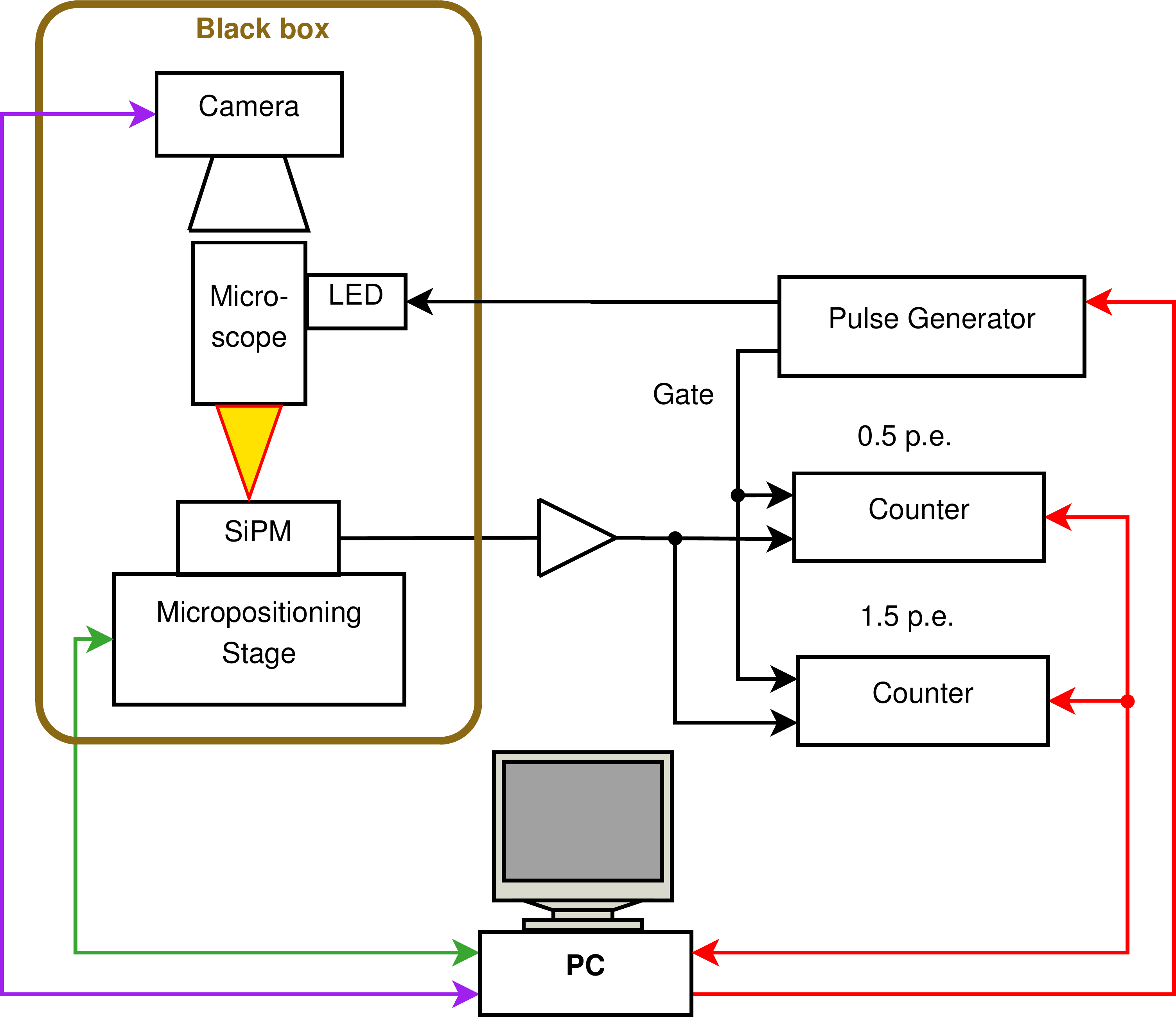}
  \caption{Test setup block scheme.}
  \label{fig:blok-schema}
\end{wrapfigure}

The measurement proceeds in following way: An LED emits a very short light pulse (in an ideal case a single photon, LED driving pulse length is \unit[10]{ns}) towards the examined sensor and the SiPM response is measured in coincidence with the electrical pulse which drives the LED. This is repeated (20~000 times) in every single spot. The whole area of a SiPM or a part of it is scanned in small steps this way, providing maps of relative PDE and crosstalk probability of the device with sub-pixel resolution.

The light source is steered through the SiPM matrix with a stepping precision of \unit[1]{\micron}. Standard commercial LEDs (dominant wavelengths \unit[470, 525 and 625]{nm}) are employed as a light source. The light from the LED passes through a small collimating aperture to a microscope (50$\times$ magnification) and is focused on surface of the sensor into a \unit[1-2]{\micron} spot. Because the microscope is equipped with a beam splitter, the surface can be observed with a CCD camera while illuminated by the light beam. With a help of this feature, we can do a coordinate calibration, which  assures focus of the light spot in any position within the SiPM plane.

The amplified output signal from the SiPM is split into two and each of those is discriminated at different voltage levels set to 0.5 and 1.5~photon equivalents (p.e.) The discriminated signals are fed into two counters, one for each branch of the original signal. The counters simply measure the number of trigger events coming out of the SiPM in coincidence with the pulse generator, which supplies also the sensor illuminating LED. The measurement at two different discrimination level allows to correct our data for crosstalk.



\section{Data analysis} \label{sec:analysis}
As mentioned above, the scan setup provides us with two data sets, maps of the SiPM recorded at 0.5 and 1.5~p.e. First of all, the noise is subtracted in both maps and the data is scaled with the number of shot light pulses to get the desired relative efficiency. The map taken at 0.5~p.e. comprises all events when one or more pixels fired, it represents the response of the detector including crosstalk. Afterpulsing does not contribute in our case, because the coincidence time window is chosen short enough to eliminate most of the afterpulsing.

The crosstalk probability is calculated by dividing the map recorded at 1.5~p.e. by the one taken at 0.5~p.e. point-by-point. In other words, the resulting map tells us the probability that more than one cell will fire, if we illuminate a particular spot on a sensor.


We can also obtain the geometrical fill-factor of the SiPM by integrating only that area, whose relative sensitivity lies above certain cut-off threshold. 

Based on the data, we can also perform a single pixel analysis. Thanks to precise alignment of the chip done via coordinate calibration, it is possible to overlay the sensor map with a rectangular grid so that each cell of the grid contains exactly one pixel of the SiPM. Since the pixels are separated, we can calculate characteristics for every single one. We portray the result as a map again. In this case, every pixel in the plot represents one real pixel on the detector. We exclude from this kind of characterization the pixels at the edge of the scan window
. We create the single pixel maps for two quantities, the relative PDE and geometrical fill-factor, both are corrected for noise and crosstalk. This feature is applicable only to SiPMs with rectangular pixels, other pixel shapes would require more complicated pixel identification algorithms.

\section{Results}
We present scan result of two SiPMs. The motivation for detailed testing of the Hamamatsu Multi-Pixel Photon Counter (MPPC) with pixel pitch \unit[50]{\micron} was its usage in the Tungsten Timing Test Beam (T3B) experiment \cite{t3b}. Since the main purpose of the setup was to test prototypes of the Silicon MultiPixel light-detector (SiMPl) devices being developed and manufactured in the Semiconductor Laboratory of the Max Planck Institute, we show one example scan of a second production series device. For the SiPMl, only the relative PDE map is presented, because the performance of the present prototypes is not optimal at room temperature (i.e. without cooling) and the crosstalk cannot be properly measured with the current equipment. Only a result for a smaller sensor is presented, because larger arrays suffer from too large dark rate. These issues will be solved in the next design iteration. A green LED was used for all measurements shown here.

\subsection{Hamamatsu MPPC}
\begin{figure}[t]
  \centering
  \includegraphics[width=0.47 \textwidth]{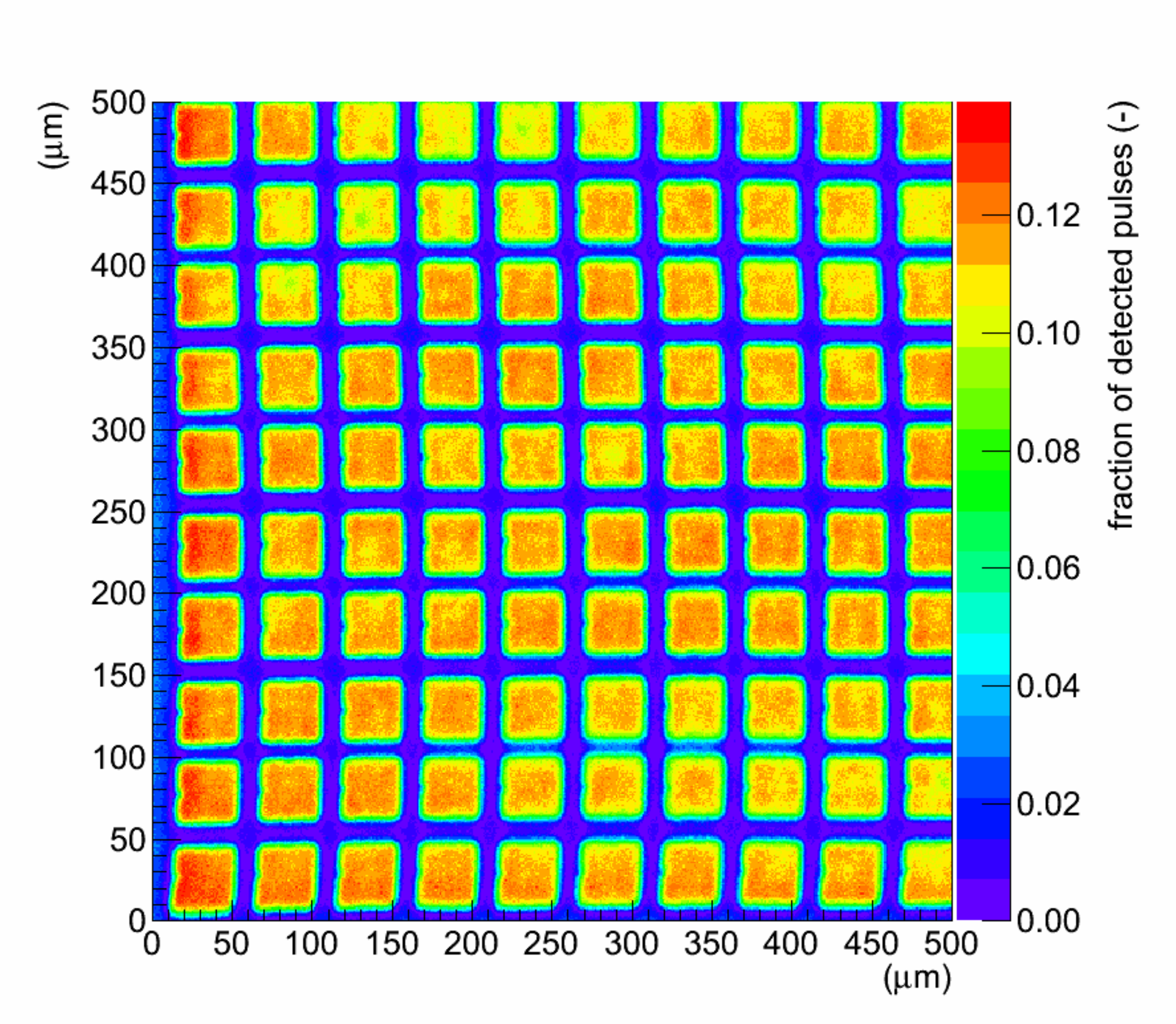}
  \hspace{0.5cm}
  \includegraphics[width=0.47 \textwidth]{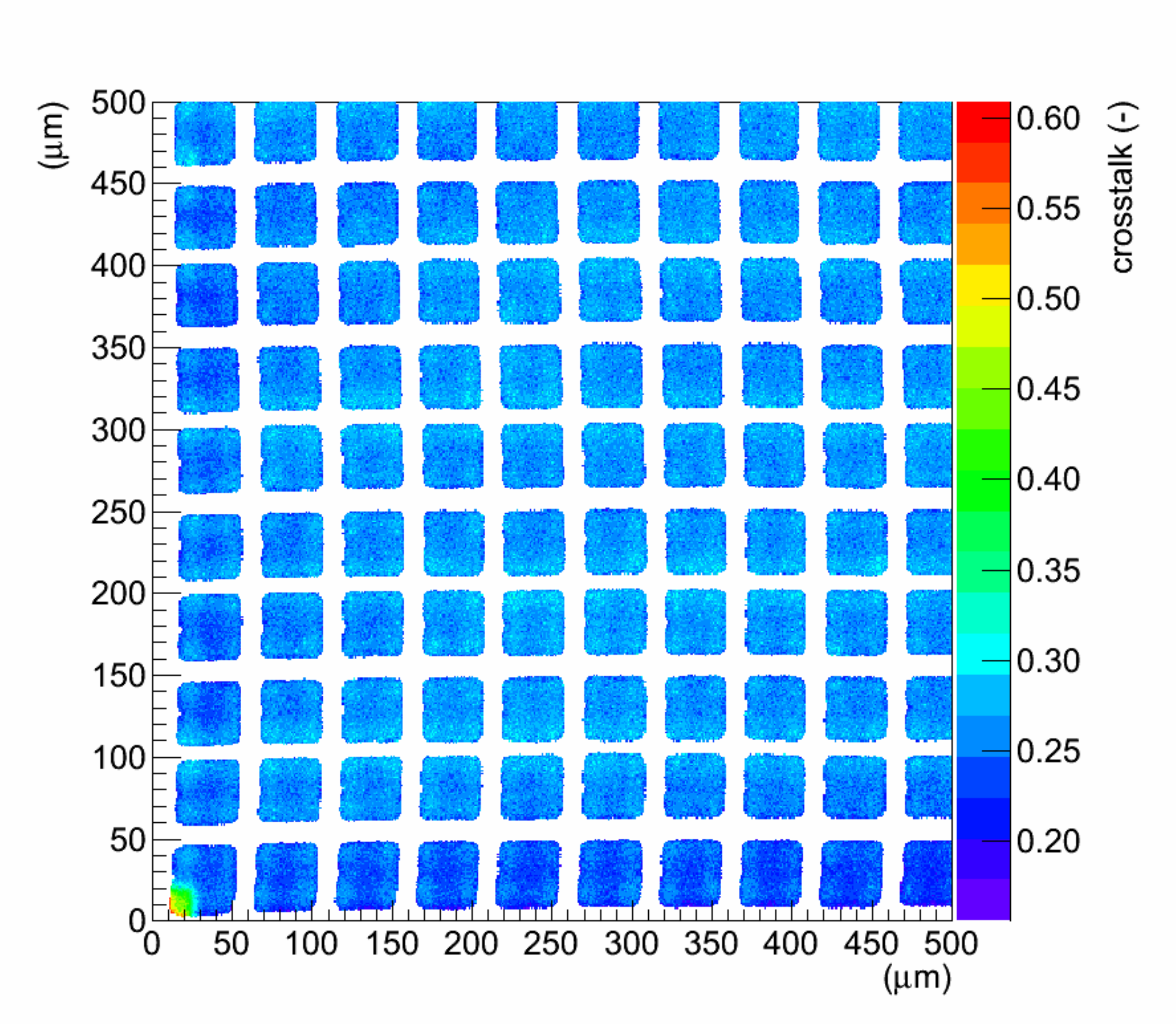}

  \caption{Maps of Hamamatsu MPPC-S10943-8584(X) (1/4 of the array). {\bf (left)} Relative PDE map. {\bf (right)} Crosstalk probability map. (Over-bias voltage 1.5~V)}
  \label{fig:Hama50meas}
\end{figure}

The relative PDE map of the Hamamatsu MPPC-S10943-8584(X) is presented in Fig.~\ref{fig:Hama50meas} left. It shows that the PDE is quite homogeneous within a pixel, except for the outermost ones. The outer pixels have a gradient in PDE heading outwards from the pixel array. However, there are apparent differences between individual pixels. The crosstalk probability map (Fig.~\ref{fig:Hama50meas} right) evinces the expected property that the crosstalk is the largest at the edge of the pixel and drops in the middle, also low crosstalk in the outermost pixels in the array was expected. The ``hot'' spot in the lower left corner is most probably a result of so called edge-breakdown. Locally higher electrical field leads to a higher gain and higher breakdown probability. This causes first a higher PDE, and second a higher signal amplitude, which exceeds the 1.5~p.e. level and hence is counted as crosstalk. The edge-breakdown can also explain the higher sensitivity of the detector edge (left and bottom edge in Fig.~\ref{fig:Hama50meas} left).

\begin{figure}[t]
  \centering
  \includegraphics[width=0.47 \textwidth]{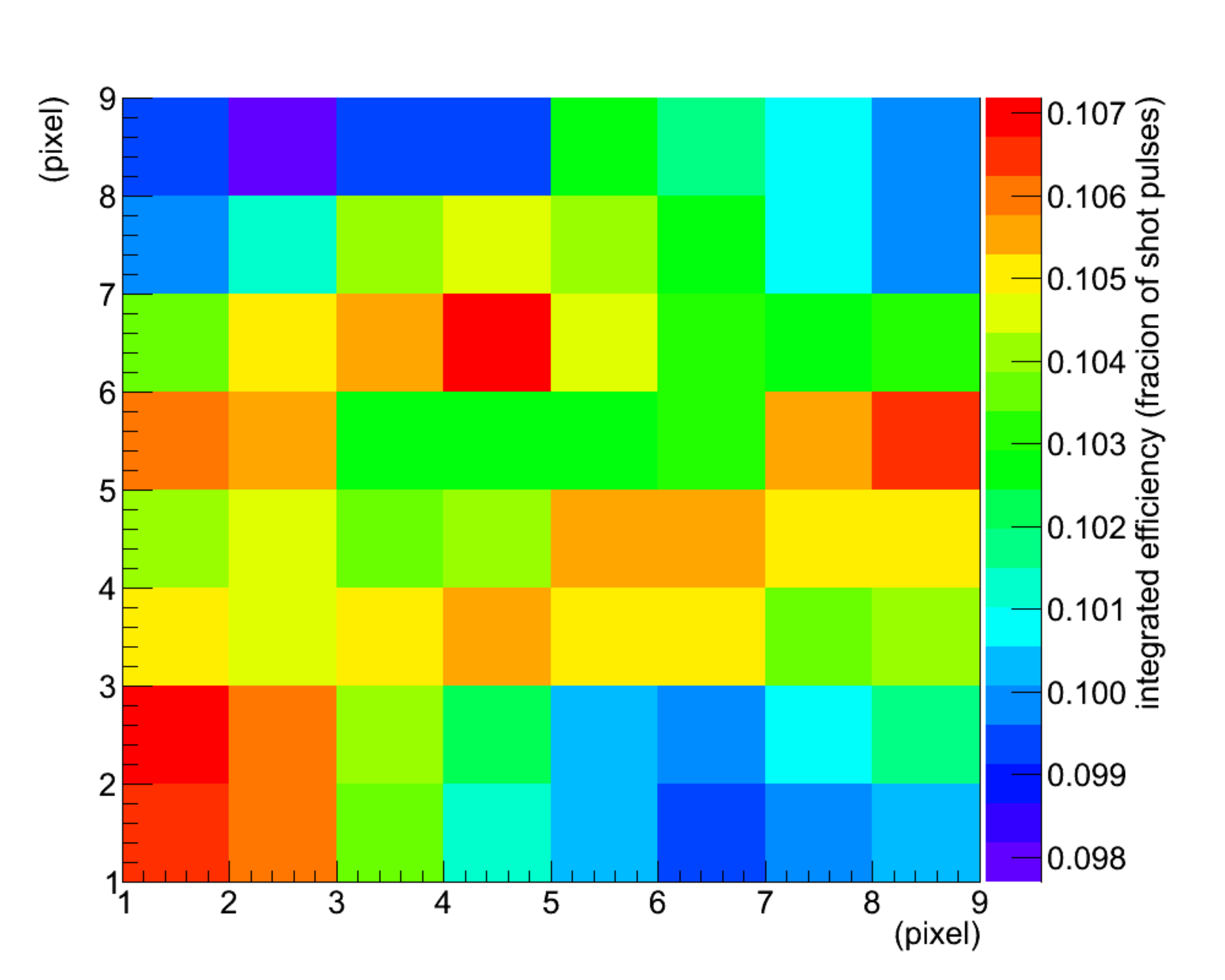}
  \hspace{0.5cm}
  \includegraphics[width=0.47 \textwidth]{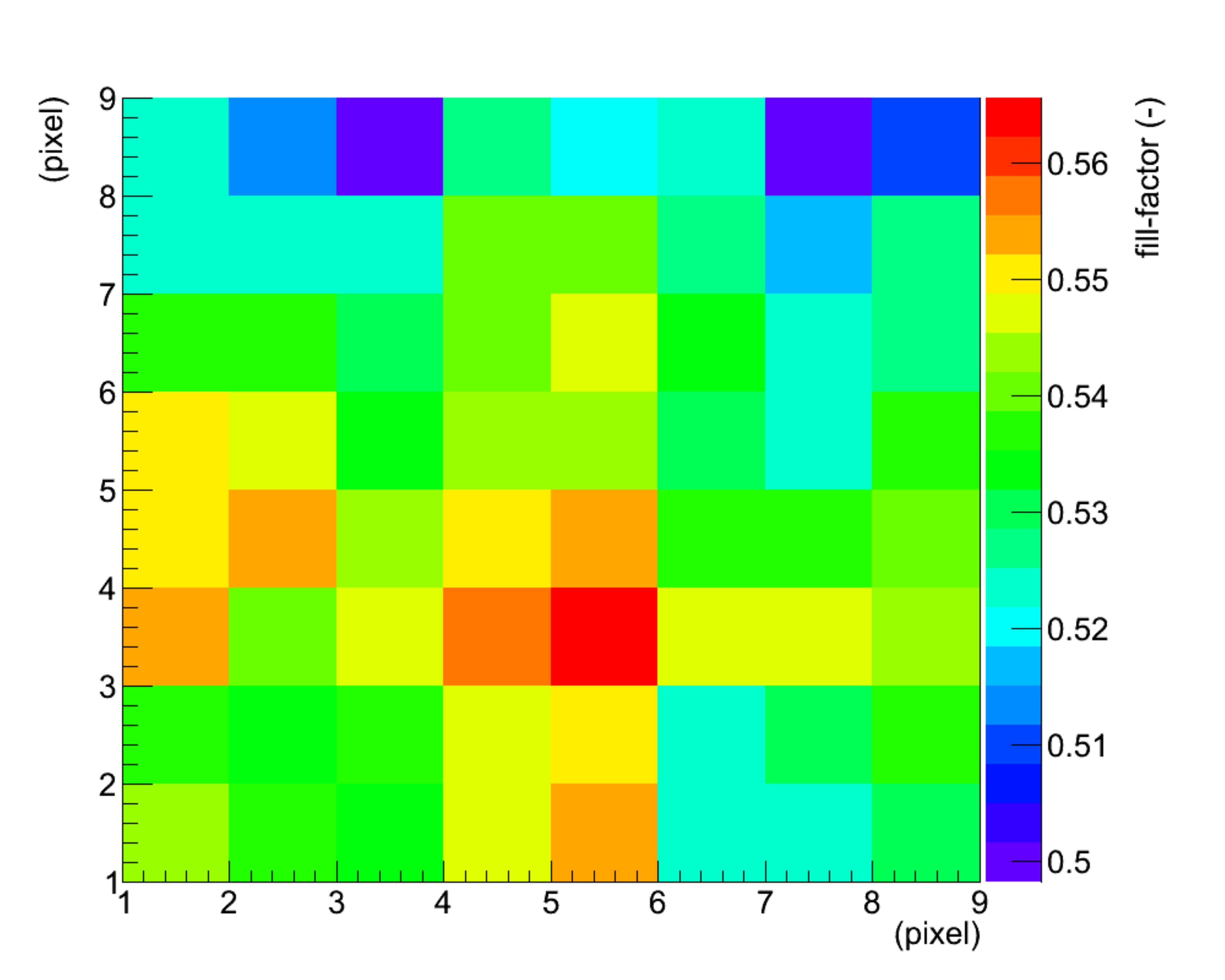}

  \caption{Sensor maps resulting from the single pixel analysis of a quarter of Hamamatsu MPPC-S10943-8584(X). {\bf (left)} Relative PDE calculated for every pixel. {\bf (right)} Geometrical fill-factor of every pixel.}
  \label{fig:Hama50sp}
\end{figure}

To quantify the PDE homogeneity across the whole sensor, we used the results of the single pixel analysis. The integrated efficiency map is shown in Fig.~\ref{fig:Hama50sp} (left) and the geometrical fill-factor map in Fig.~\ref{fig:Hama50sp} (right). Each pixel in the plot represents one pixel on the SiPM (the outermost pixels are dropped). To characterize the SiPM homogeneity we calculated two quantities, the relative spread in the PDE and in the fill-factor between the minimum and maximum on the map. These numbers are listed in Tab.~\ref{tab:H50-vysledky} along with overall geometrical fill-factor and crosstalk probability. The crosstalk value presented in this table was measured in a black box without any illumination. We obtained a higher crosstalk the measurements with light compared to the one without. The reason is presumably stray light reflected or scattered on the parts of the setup or in the SiPM plastic casing.

We also examined changes of patterns on all maps if we use different bias voltage. We conducted the same scan for three different bias voltages. We did not observe any significant changes in the patterns in plots corresponding to Figs.~\ref{fig:Hama50meas} and \ref{fig:Hama50sp}. Only the crosstalk map evinced substantial changes, the effect of the edge breakdown was reduced with growing bias voltage. The overall geometrical fill-factor stayed unaffected.

\subsection{SiMPl device}
As discussed above, with the current SiMPl \cite{nin-hQE} prototypes, a full characterization is not yet possible. Here, we show one proof-of-principle result, a scan of a so called ``double flower'' structure (19 pixels) with a pitch of \unit[135]{\micron} and \unit[13]{\micron} gap between neighboring pixels (see Fig.~\ref{fig:simpl}). This measurement confirms that the shape of the sensitive area and fill-factor matches the design, and shows validity of the detector concept. It is apparent that the inner pixels have about 20\% higher response to light compared to the outer ones. This can be an effect of crosstalk, which is not subtracted here, or it can be caused by slightly different design of the outermost pixels, or both. Also the effect of the edge breakdown can be seen very weakly, it is visible in the inner pixels at the upper left edge.

\section{Summary}
In this publication, we presented a part of an extensive study of a Hamamatsu MPPC (\unit[50]{\micron pitch}) and a proof-of-principle test of a SiMPl device. This was achieved with use of a newly built measurement setup.

\begin{minipage}{\textwidth}
  \begin{minipage}[b]{0.47 \textwidth}
    \centering
    \begin{tabular}{cc}
      {\bf Quantity}		& {\bf Value} \\
      \midrule
      PDE spread		& 8\% \\
      Fill-factor spread	& 11\% \\
      Crosstalk probability	& 18\% \\
      Geometrical fill-factor	& 55\% \\
      \bottomrule
      \vspace{4ex}
    \end{tabular} 
    \captionof{table}{Study results for Hamamatsu MPPC-S10943-8584(X).}
    \label{tab:H50-vysledky}
  \end{minipage}
  \hfill
  \begin{minipage}[b]{0.47 \textwidth}
    \centering
    \includegraphics[width=0.95 \textwidth]{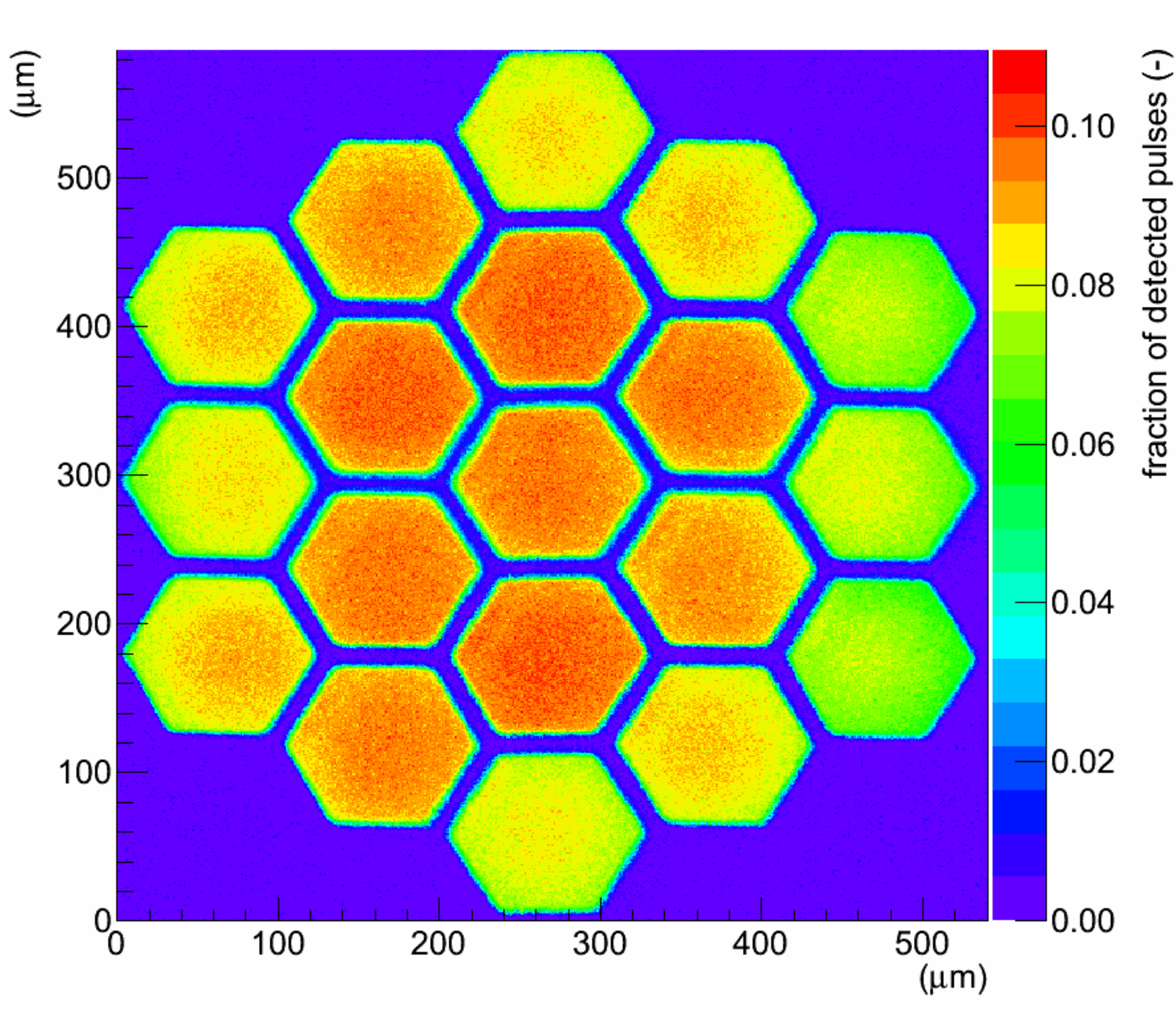}
    \captionof{figure}{Sensitivity map of SiMPl (P13) ``double flower'' structure.}
    \label{fig:simpl}
  \end{minipage}
\end{minipage}
\vspace{3ex}

The Hamamatsu MPPC study showed that the response of the sensor is slightly higher at the edges of the pixels, most probably because of inhomogeneous electrical field across the pixel of changing doping profiles. This effect is even more pronounced in the outermost pixels. It was observed that the patterns in PDE and in fill-factor distributions for single pixels do not significantly change with bias voltage. The overall geometrical fill-factor also was not affected, that means that the sensitive area size is voltage independent. Our observations are consistent with \cite{tadday}.

The test of a \unit[135]{\micron} pitch SiMPl device proved that the shape of the active area of these detectors has the expected form and that the sensor with this design can operate successfully. Some issues regarding PDE gradients across the pixel matrix have been discovered. All measurements were conducted at room temperature which is not favorable for the presently available SiMPls and therefore not all the tests could have been done. The performance of the SiMPls at room temperature will be improved in the next iteration of these devices.



\end{document}